\documentclass[onecolumn]{aastex631}
\usepackage{textcomp}
\usepackage{graphicx}
\usepackage{amssymb}
\usepackage{amsmath}
\usepackage{epstopdf}
\usepackage{gensymb}
\usepackage{natbib}
\usepackage{appendix}
\usepackage{tabu}
\usepackage{color}
\usepackage{multirow}
\usepackage{booktabs}
\usepackage{ulem}
\usepackage{threeparttable}
\definecolor{darkblue}{rgb}{0.17, 0.49, 0.72}
\definecolor{darkgreen}{rgb}{0.0, 0.4, 0.0}
\newcommand{\sref}[1]{Section~\ref{#1}}
\newcommand{\fref}[1]{Figure~\ref{#1}}
\newcommand{\eqn}[1]{Equation~(\ref{#1})}

\defcitealias{Sowmyaetal2021}{Paper I}
\begin{document}
\title{Solar variability in the Mg II h and k lines}
%\received{\today}
%\acceptjournal{ApJ}
\shortauthors{Sowmya et al.}

\correspondingauthor{K.~Sowmya}
\email{krishnamurthy@mps.mpg.de}

\author[0000-0002-3243-1230]{K.~Sowmya}
\affiliation{Max-Planck-Institut f\"ur Sonnensystemforschung, Justus-von-Liebig-Weg 3, 37077 G\"ottingen, Germany}
\affiliation{Institute of Physics, University of Graz, Universit\"atsplatz 5, 8010 Graz, Austria}

\author[0000-0001-9106-1332]{M.~Snow}
\affiliation{Laboratory for Atmospheric and Space Physics, University of Colorado Boulder, 1234 Innovation Dr., Boulder CO 80303, USA}
\affiliation{South African National Space Agency, Hospital Street, Hermanus 7200, South Africa}
\affiliation{Department of Physics and Astronomy, University of the Western Cape, Robert Sobukwe Road, Bellville 7535, South Africa}

\author[0000-0002-8842-5403]{A.~I.~Shapiro}
\affiliation{Institute of Physics, University of Graz, Universit\"atsplatz 5, 8010 Graz, Austria}
\affiliation{Max-Planck-Institut f\"ur Sonnensystemforschung, Justus-von-Liebig-Weg 3, 37077 G\"ottingen, Germany}

\author[0000-0002-1377-3067]{N.~A.~Krivova}
\affiliation{Max-Planck-Institut f\"ur Sonnensystemforschung, Justus-von-Liebig-Weg 3, 37077 G\"ottingen, Germany}

\author[0000-0002-0335-9831]{T.~Chatzistergos}
\affiliation{Max-Planck-Institut f\"ur Sonnensystemforschung, Justus-von-Liebig-Weg 3, 37077 G\"ottingen, Germany}

\author[0000-0002-3418-8449]{S.~K.~Solanki}
\affiliation{Max-Planck-Institut f\"ur Sonnensystemforschung, Justus-von-Liebig-Weg 3, 37077 G\"ottingen, Germany}

%TC:ignore
\begin{abstract}
Solar irradiance and its variations in the ultraviolet (UV) control the photochemistry in Earth's atmosphere and influence Earth's climate. The variability of Mg II h and k core-to-wing ratio, also known as the Mg II index, is highly correlated with the solar UV irradiance variability. Because of this, Mg II index is routinely used as a proxy for solar UV irradiance variability, which can help to get insights into the influence of solar UV irradiance variability on Earth's climate. Measurements of the Mg II index, however, have only been carried out since 1978 and do not cover the climate relevant timescales longer than a few decades. Here we present a model to calculate the Mg II index and its variability based on the well-established SATIRE (Spectral And Total Irradiance REconstruction) model. We demonstrate that our model calculations yield an excellent agreement with the observed Mg II index variations, both on the solar activity cycle and on the solar rotation timescales. Using this model, we synthesize Mg II index timeseries on climate relevant timescales of decades and longer. Here we present the timeseries of the Mg II index spanning nearly three centuries.
\end{abstract}
%TC:endignore

\keywords{Solar activity (1475) -- Solar ultraviolet emission(1533) -- Solar chromosphere(1479) -- Plages (1240) -- Solar transition region(1532)}

\section{Introduction}
\label{sec:intro}
The solar radiation, which is the main source of energy input to the Earth, varies at all measurable timescales and wavelengths. The constantly evolving solar surface magnetic fields are established to be the main drivers of such variations on timescales longer than a day \citep{Krivovaetal2003, KL2017, Shapiroetal2017}. These evolving surface magnetic fields extend to the chromosphere, where the bulk of the ultraviolet (UV) radiation forms. There they lead to local brightenings that regulate the solar UV output. Variations in the solar UV radiation influence Earth's climate via the so called top-down mechanism \citep[see e.g. reviews by][]{Grayetal2010,Ermollietal2013,Solankietal2013}. An accurate reconstruction of the solar UV irradiance and its variability is therefore crucial for understanding the role of solar variability in causing natural climate change. 

Due to the lack of long and stable, directly measured timeseries of solar UV irradiance, studies of the impact of solar UV irradiance on Earth's climate often rely on proxies. The most important of solar UV irradiance proxies is the Mg II core-to-wing ratio that is also often called the Mg II index \citep[][see also \citealt{Snowetal2019}]{HeathandSchlesinger1986} based on the Mg II h and k emission around 280\,nm. Because it is a ratio of fluxes (at nearby wavelengths), it is less  prone to instrumental effects than the fluxes themselves. Further, the Mg II index is known to be strongly correlated with the solar UV irradiance below 400\,nm \citep[e.g.][]{HeathandSchlesinger1986,ViereckandPuga1999,Vierecketal2001} and it has even been used as a direct input in Earth's atmosphere modeling \citep[e.g.][see also \citealt{ThuillierandBruinsma2001,Bruinsmaetal2003}]{Thuillieretal2012}. However, the measurements of this index span only a few decades (starting from 1978) and therefore different approaches have been developed to create a continuous and longer timeseries of Mg II index which is required for Earth's atmosphere and climate studies. 

A number of reconstructions of Mg II index data are based on the observed correlation between Mg II index and other solar indices. For example, \citet{DelandandCebula1993} used the correlation between Mg II index and the solar radio flux at 10.7\,cm to create a composite of Mg II index covering the period 1978-1992. By designing a model for the Mg II index data from 10.7\,cm flux using neural networks, \citet{Tebabaletal2017} extrapolated the Mg II index to 1947. Recently, \citet{Royetal2021} exploited the correlation between Mg II index and 10.7\,cm to fill gaps in existing Mg II index observations. Further, Mg II h and k emission shows a strong correlation with Ca II H and K emission \citep{Donnellyetal1994} since they form around similar heights in the solar atmosphere and trace the same plasma \citep[][see also \citealt{Linsky2017}]{Vernazzaetal1981}. Based on this property, \citet{Morrilletal2011} developed an approach using Ca II K images from Mt. Wilson observatory to estimate Mg II index between 1961 and 1981. Using the correlations found between the Mg II index and neutron-monitor data and applying them to the cosmogenic isotope data, \citet{Thuillieretal2012} reconstructed Mg II index back to the Maunder minimum period. 
 
Another class of approaches is based on combining synthetic spectra of magnetic features and quiet-Sun with observed distribution of magnetic features on the solar disk. For example, \citet{Criscuolietal2018} used synthetic spectra and area coverages of the magnetic features determined using Ca II K and photospheric red continua full disk images from several ground-based telescopes to reconstruct Mg II index back to 1989. It was later extended further back to 1749 by \citet{Berrillietal2020}. 

Here we take a similar approach but build on the success of the SATIRE-S (Spectral And Total Irradiance REconstruction for the Satellite era) model which has a very comprehensive methodology for defining the disk distribution of magnetic features. It directly uses magnetic field information from observed full disk solar magnetograms, and intensity images to determine the area coverage of magnetic features \citep{Fliggeetal2000,Krivovaetal2003,Krivova2011,Yeoetal2014}. This allowed SATIRE-S model to yield a remarkable agreement with observations of total and spectral solar irradiance. 

Traditionally SATIRE-S model employed spectra from \citet{Unruhetal1999} synthesized in local thermodynamic equilibrium (LTE) using 1D atmospheres without a chromosphere. This is sufficient for calculating solar irradiance longward of 300\,nm but requires an empirical correction below it \citep{Krivovaetal2006, Yeoetal2014}. Recently \cite{Tagirovetal2019} included non-LTE modelling into SATIRE-S and reconstructed solar UV irradiance (with 1 nm spectral resolution, not resolving strong spectral lines such as Ca II H and K, Mg II h and k, and hydrogen lines) circumventing the need for the empirical correction. Later \cite{Sowmyaetal2021} used SATIRE-based approach  to model the variability of solar Ca II H and K lines. Here we focus  on modelling the high resolution spectral profile of Mg II h and k lines. Namely, we take SATIRE-S distributions of magnetic features and combine them with Mg II h and k spectra of quiet-Sun and magnetic features. These spectra are obtained using  1D semi-empirical solar atmospheric models and include a proper treatment of the physical mechanisms governing the properties of Mg II h and k lines. In particular, they account for non-LTE effects and partial frequency redistribution in line cores.

Our approach is described in \sref{sec:model}. In \sref{sec:resobs}, we show that the synthetic Mg II index values are in excellent agreement with the observations from SOLar-STellar Irradiance Comparison Experiment (SOLSTICE) on the SOlar Radiation and Climate Experiment \citep[SORCE;][]{Rottman2005}. We further demonstrate that our approach reproduces the observed variations of Mg II index both on the solar rotation and activity cycle timescales. The extension of the Mg II index timeseries to 1745 is described in \sref{sec:resopenflux} and conclusions are presented in \sref{sec:concl}.

\section{Methods}
\label{sec:model}
Our SATIRE-S based model attributes irradiance changes on timescales longer than a day to the changes in the surface magnetic field, which manifests as dark sunspots and bright faculae (although the chromospheric counterparts of faculae are called plages, we refer to both faculae and plages collectively as `faculae' throughout the paper for simplicity). The disk integrated solar flux $F$ at a given time $t$ and wavelength $\lambda$ is therefore computed by combining the area coverages of quiet-Sun (regions free of the magnetic features), sunspots and faculae with their corresponding brightness. However, we neglect sunspots following the finding that solar UV emission is dominated by contribution from faculae \citep{LeanandRepoff1987,Leanetal1997,Unruhetal2008,Penzaetal2023,Sowmyaetal2023}. The quiet-Sun and facular fluxes are computed by decomposing the solar disk into $l$ concentric rings whose positions are defined by $\mu_l$ (cosine of the heliocentric angle $\theta$). Namely,

\begin{equation}
\begin{split}
    F(t,\lambda) & = \sum_{l} I_q(\lambda,\mu_l)\Delta\Omega_l+ \sum_{l} \alpha_{fl}(t)\left[I_f(\lambda,\mu_l)-I_q(\lambda,\mu_l)\right]\ \Delta\Omega_l\ ,
\end{split}
    \label{eq:ssi}
\end{equation}
where $\alpha_{fl} (t)$ are the time dependent fractional area coverages of faculae in the $l$th ring which subtends a solid angle of $\Delta\Omega_l$ at a distance of 1\,AU from the Sun. $I_q(\lambda,\mu_l)$ and $I_f(\lambda,\mu_l)$ are the specific intensities of quiet-Sun and faculae, respectively.

The specific intensities in \eqn{eq:ssi} are computed using the RH code \citep{Uitenbroek2001} by including the non-LTE and partial frequency redistribution effects which are essential to explain the shapes of the Mg II h and k lines \citep[see e.g.][]{Uitenbroek1997}. A 22-level configuration for the Mg atom provided in the RH code is used (MgI\_b+MgII.atom). The atmospheric structure of the quiet-Sun is represented by model C while that of faculae by model P of \citet{Fontenlaetal1999}. Background lines in the wavelength window between 274 and 286\,nm are included from Kurucz's line list\footnote{\url{http://kurucz.harvard.edu/linelists.html}} to determine background line opacity and emissivity in LTE conditions. 

To validate the modeled spectra synthesized using the approach above, we use level 2 quiet-Sun disk center raster observations from Interface Region Imaging Spectrograph \citep[IRIS;][]{DePontieuetal2014} and Total and Spectral Solar Irradiance Sensor-1 (TSIS-1) hybrid solar reference spectrum from \citet{Coddingtonetal2023}. IRIS observations used here were recorded on 05 October 2022 between 04:10:19 UT and 04:28:24 UT, and cover a field of view of 127''x119''. IRIS radiometric calibration procedure on SSW IDL was used to convert the data number units to intensity in physical units (W m$^{-2}$ Sr$^{-1}$ nm$^{-1})$.

\begin{figure*}
    \centering
    \includegraphics[scale=0.7]{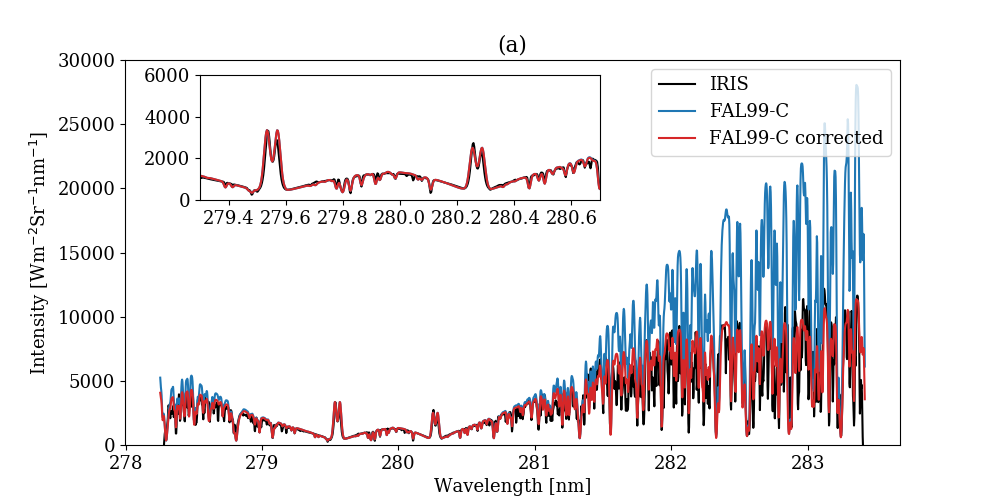}
    \includegraphics[scale=0.7]{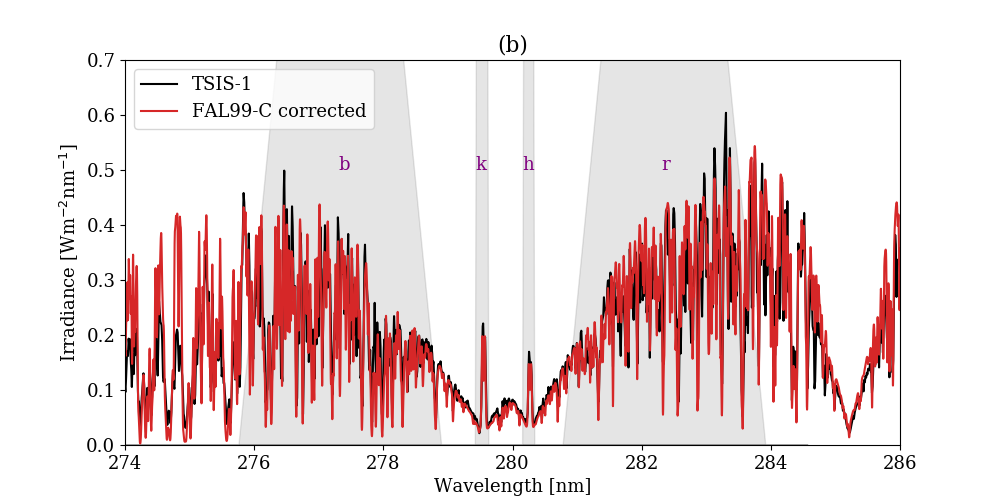}
    %IRIS spectra was observed on 05 October 2022
    \caption{Comparison between observed and modeled spectra. Panel (a): disk-center quiet-Sun spectrum observed by IRIS (black), FAL99-C spectrum before (blue) and after (red) applying the H-minus opacity corrections in spectral synthesis. Panel (b): quiet-Sun irradiance spectrum from \citet[][black]{Coddingtonetal2023} and FAL99-C after opacity corrections (red). Gray shaded areas marked b, k, h, and r are the masks used to determine the Mg II index. See \sref{sec:model} for details.}
    \label{fig:obscal}
\end{figure*}

Our synthetic quiet-Sun spectrum reproduces the line core and near-wing features observed by IRIS remarkably well as shown in \fref{fig:obscal}a (see also inset). However, the synthetic spectrum overestimates the intensities in the far-wings of the Mg II lines (compare spectra shown in blue and black in \fref{fig:obscal}a). This is due to the well known missing opacity issue in the UV arising from incomplete spectral line lists \citep{Kurucz2009}. To correct for these missing spectral lines, we followed the  H-minus opacity correction approach introduced by \citet{Brulsetal1992}. The disk-center spectrum obtained after the opacity correction is shown in red in \fref{fig:obscal}a. For simplicity, the correction factors are assumed to be independent of $\mu_l$. Our synthetic disk-integrated spectral flux, computed under such an assumption, results in a good agreement with the TSIS-1 spectrum, as can be seen in \fref{fig:obscal}b. We note that line broadening by macroturbulent velocity fields is taken into account by convolving the Mg II line cores with a Gaussian profile having FWHM of 200\,m\AA{} (corresponding to a velocity of $\sim 21$\,km\,s$^{-1}$). Spectral regions outside the line cores, which form lower in the atmosphere than the line cores (mainly the photospheric spectral lines in the Mg II line wings),  are convolved with a Gaussian profile of 100\,m\AA{} (corresponding to a velocity of $\sim 10$\,km\,s$^{-1}$). These velocities, required to reproduce the observed line widths, are consistent with the plasma velocity amplitudes in the formation layers of Mg II lines, as determined from 3D magnetohydrodynamic simulations \citep{Ondratscheketal2024}.

The fractional area coverages of faculae in \eqn{eq:ssi}, for the time period 1999--present, are taken from \citet{Yeoetal2014} and extended further to the present time with their method \citep[][Yeo, K. L., private communication]{Sowmyaetal2021}. The method of \citet{Yeoetal2014} is based on the photospheric full disk continuum intensity and magnetogram images from ground- and space-based observations. Although the facular area coverages recovered with their approach goes back to 1975 (see \fref{fig:mglong}), we limit it to 1999 (a) due to the gaps in the facular area coverages for the solar cycles before 1999 and (b) due to the availability of better quality space-based solar continuum intensity and magnetogram images starting from 1999 (basically uninterrupted after the SOHO "vacation"). For the reconstruction of Mg II indices in the interval 1745--1998, the fractional area coverages of faculae from \citet[][see also \citealt{Natasha2007}]{Wuetal2018} are used. \citet{Wuetal2018} reconstructed the past evolution of the solar magnetic flux from the sunspot number, which is the only available direct solar activity metric before the late 19th century, by solving a set of coupled ordinary differential equations \citep{Solankietal2000,Solankietal2002,VieiraandSolanki2010}. The magnetic flux was then converted into fractional area coverages of the magnetic features as described by \citet{Krivovaetal2006}.

For calculating the Mg II index, we follow a new definition presented in \citet{Snowetal2019} based on SORCE/SOLSTICE spectra at 0.1\,nm resolution. This revised definition uses weighted sums of fluxes at specific spectral regions in core and wing (shown as gray shaded b, h, k, and r regions in \fref{fig:obscal}b) instead of using fluxes at 7 specific wavelength positions as in the traditional definition \citep{HeathandSchlesinger1986}. We refer to \citet{Snowetal2019} for a detailed discussion of the need for this new algorithm to calculate Mg II index. To be consistent with definition of \citet{Snowetal2019}, we smooth the synthetic spectra to a spectral resolution of 0.1\,nm \citep{Snowetal2005b} and resample them to the SORCE/SOLSTICE wavelength grid with a spacing of 0.02\,nm. We then compute the Mg II index and its time variations using:
\begin{equation}
    {\rm Mg\ II\ index} = \frac{C_{\rm k}+C_{\rm h}}{W_{\rm b}+W_{\rm r}}\ ,
    \label{eq:mgindex}
\end{equation}
where $C_{\rm k}$ and $C_{\rm h}$ are the weighted sums of fluxes in the cores while $W_{\rm b}$ and $W_{\rm r}$ are the weighted sums of fluxes in the wings. They are obtained as
\begin{equation}
    \frac{\sum_{b,h,k,r} x_{\lambda} * F(t,\lambda)}{\sum_{b,h,k,r} x_{\lambda}}\ , \nonumber
\end{equation}
where $x_{\lambda}$ are the wavelength-dependent weights from \cite{Snowetal2019}. While calculating the flux in the line core regions, we multiply the facular area coverages by a constant factor following \citet{Sowmyaetal2021}. This accounts for the expansion of flux tubes forming faculae as they rise to the chromosphere where the Mg II h and k line cores form. By carrying out a linear regression of the calculated Mg II index values with those observed, we determine the value of this expansion factor to be 3.0. This value of the expansion factor gives a regression slope very close to unity. It is reassuring that this value is close to the expansion factor of 2.9 used by \citet{Sowmyaetal2023} to model Ca II H and K emission, since the Mg II h and k emission forms around similar heights.

\begin{figure*}[ht!]
    \centering
    \includegraphics[scale=0.7]{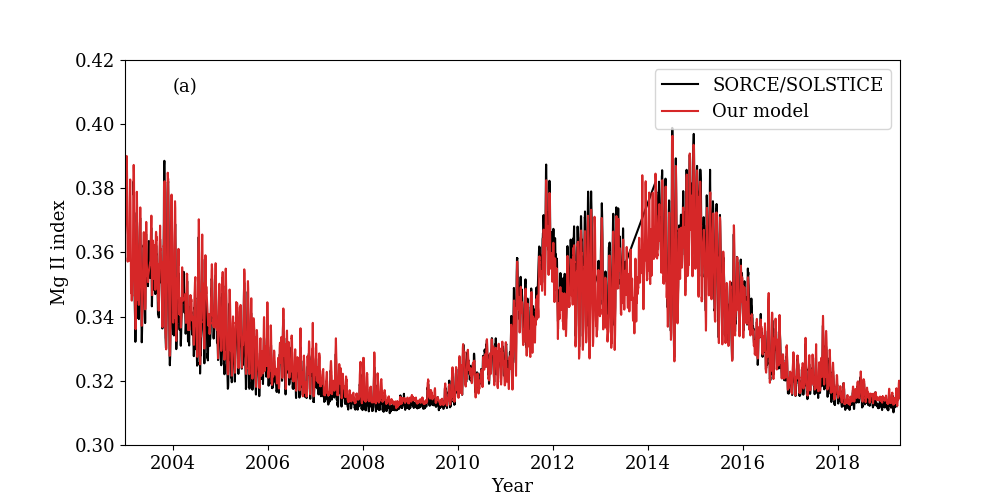}
    %\vskip -1cm
    \includegraphics[scale=0.7]{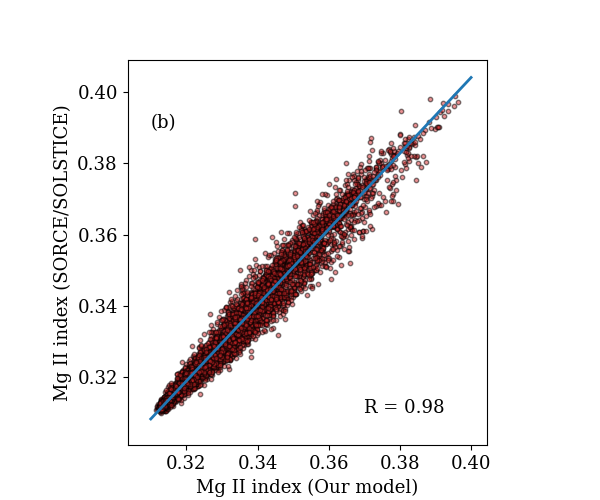}
    \caption{Comparison between observed and modeled Mg II indices. Panel (a): daily values from SORCE/SOLSTICE (black) and our model (red). Panel (b): linear regression of the observed and modeled Mg II indices (performed by choosing only those days for which both observed and modeled data are available i.e. only common available days). The Pearson linear correlation coefficient ${\rm R}=0.98$.}
    \label{fig:mgindex}
\end{figure*}

\begin{figure*}
\centering
\includegraphics[scale=0.7]{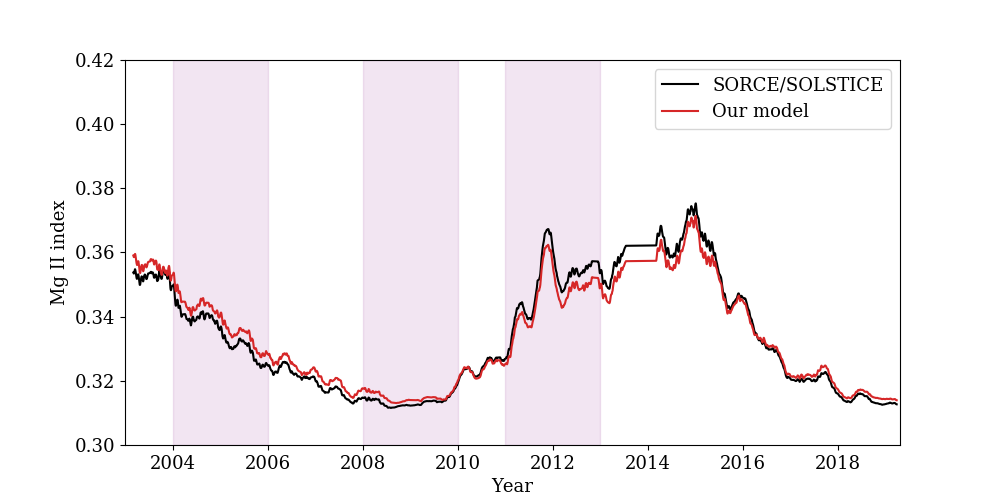}
\caption{Solar cycle variability of the Mg II index, calculated by taking 81-day running averages of the Mg II indices shown in \fref{fig:mgindex}. Observations from SORCE/SOLSTICE are shown in black and model calculations are shown in red. The shaded regions mark the intervals chosen to investigate rotational variability (see \fref{fig:mgrot}).}
\label{fig:mgcyclic}
\end{figure*}

\section{Results}
\subsection{Comparison with SORCE/SOLSTICE measurements}
\label{sec:resobs}
To test the performance of our model, we compute Mg II indices using the facular area coverages obtained following \citet{Yeoetal2014} and compare them with SORCE/SOLSTICE Mg II index measurements from \citet{Snowetal2019}. \fref{fig:mgindex}a shows a comparison of the timeseries of daily Mg II indices from our model with the data while \fref{fig:mgindex}b shows the correlation between the two records on common available days. The high correlation between the observations and the model (0.98) is reassuring.

For a better assessment of the agreement between our calculations and observations we separately consider Mg II index variability on the activity cycle and on the solar rotation timescales. Our calculations reproduce variability on the activity timescale reasonably well. This is showcased in \fref{fig:mgcyclic} where we plot 81-day (corresponding to three solar rotations) running averages of observed and modelled Mg II indices. However, \fref{fig:mgcyclic} also points to some systematic deviations. The model values are consistently higher than what is observed prior to 2010. Between 2010 and 2016, the model values are lower than observed while beyond 2016, the model values switch again to being slightly higher than observations. These deviations could either reflect inaccuracies in the model (for example, in facular area determinations or in spectral synthesis) or instrumental effects.

The deviations could possibly have an instrumental origin, as the SOLSTICE data might be affected by calibration issues over some periods of time. \citet{Woodsetal2021} describe the operational challenges to the SORCE spacecraft throughout the mission. The primary issues affecting the Mg II measurements are the loss of stellar calibration observations from 2008 onwards and the two extended spacecraft safe holds in 2011 and 2013 due to degradation of the spacecraft battery system \citep{Snowetal2022}. Significant calibration corrections needed to be applied after the 2013 safe hold which lasted for eight months until early 2014. These wavelength-dependent calibration corrections may indeed have produced a drift in the Mg II index.
 
\begin{figure*}
    \centering
    \includegraphics[scale=0.7]{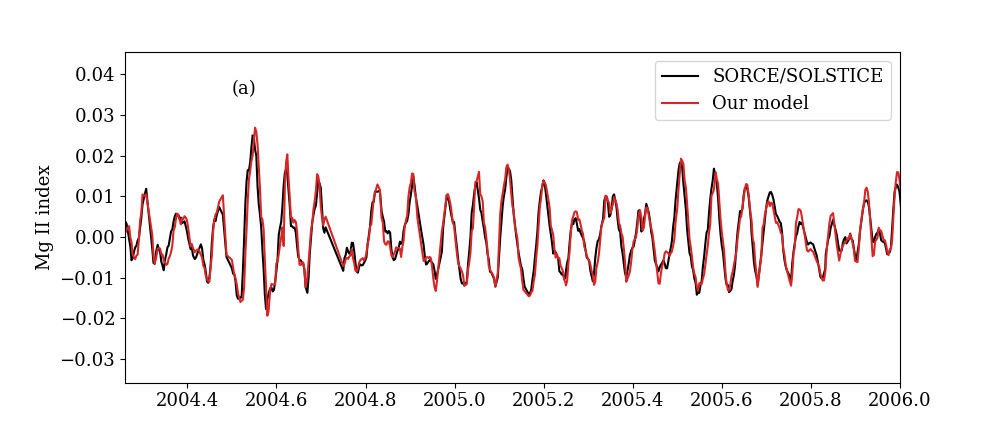}
    \vskip -0.8cm
    \includegraphics[scale=0.7]{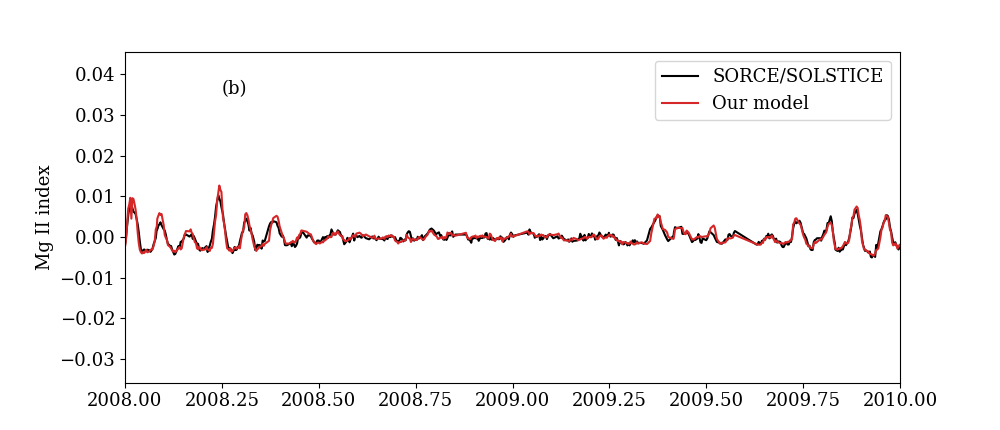}
    \vskip -0.8cm
    \includegraphics[scale=0.7]{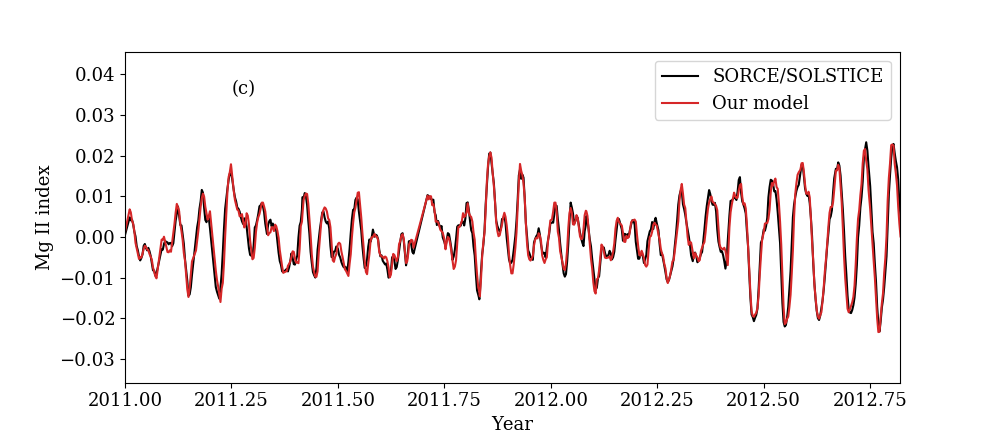}
    \caption{Rotational variability of the Mg II index during the declining phase of cycle 23 (panel a), during activity minimum of cycles 23--24 (panel b), and  close to the cycle 24 maximum (panel c). Model calculations are shown in red while observations are in black.}
    \label{fig:mgrot}
\end{figure*}

Rotational variability is obtained by subtracting the 81-day running averages from the daily Mg II index timeseries. \fref{fig:mgrot} compares the modeled variability over the 27-day rotation period with observed variations at three time intervals indicated by the red shaded regions in \fref{fig:mgcyclic}. These intervals correspond to the time when modeled Mg II index values on activity cycle timescale are higher than values observed by SORCE/SOLSTICE (\fref{fig:mgrot}a), to the cycle minimum period when the two are comparable (\fref{fig:mgrot}b), and to the time when model values are lower than observations (\fref{fig:mgrot}c). The agreement between the model and observations is remarkable irrespective of the level and phase of the activity cycle.  

\subsection{Mg II index composite timeseries}
\label{sec:resopenflux}
\begin{figure*}
    \centering
    \includegraphics[scale=0.7]{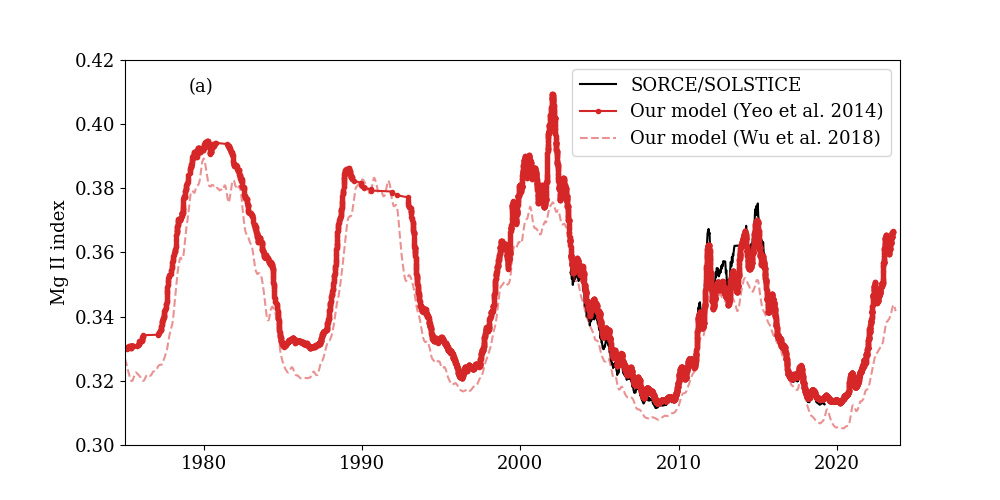}
    \vskip -1cm
    \includegraphics[scale=0.7]{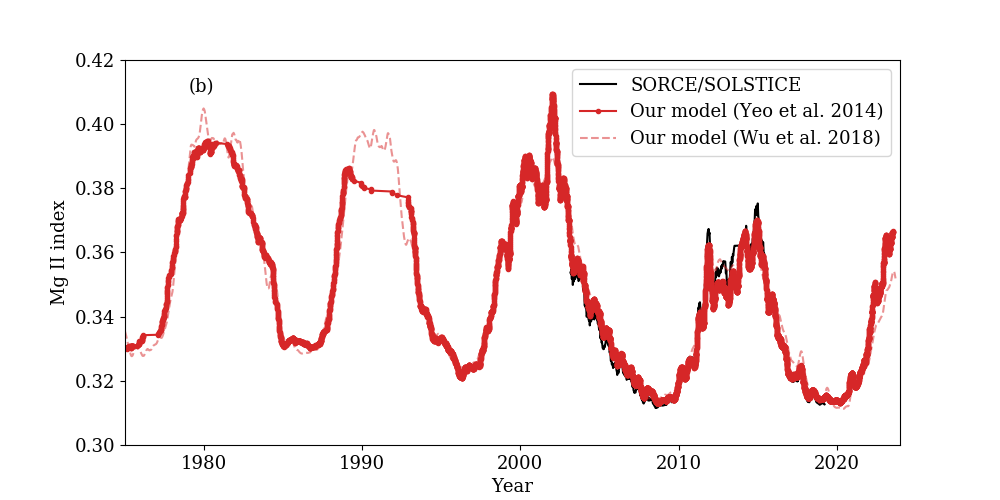}
    \caption{Comparison between 81-day running averages of the Mg II indices calculated using two different facular area inputs. Red dashed line: with facular area coverages based on \citet{Wuetal2018}, red solid line: with facular area coverages based on \cite{Yeoetal2014}. For completeness, the observations from SORCE/SOLSTICE are also shown (black). Panel (a) shows the two original calculated timeseries while panel (b) shows the timeseries after offset correction. See \sref{sec:resopenflux} for details.} 
    \label{fig:mglong}
\end{figure*}

Since the calculated timeseries presented in \sref{sec:resobs} based on \cite{Yeoetal2014} can only be extended as far back as 1975, we reconstruct Mg II indices back to 1745 using the facular areas from \citet{Wuetal2018}. A glimpse of this reconstructed Mg II index timeseries, covering the period between 1975 and the present, is shown by the red dashed lines in \fref{fig:mglong}a. By design (essentially owing to the fact that only sunspot observations are available over such long periods of time), these reconstructed facular coverages are only accurate on time scales longer than the solar rotation. Therefore, we use 81-day running averages of the modeled Mg II index values to prepare the composite timeseries. 

\begin{figure*}
    \centering
    \includegraphics[scale=0.7]{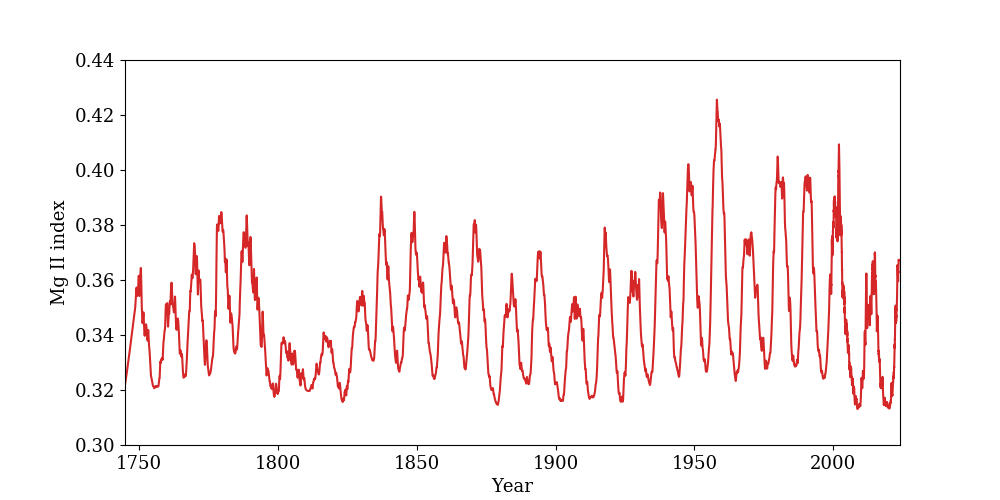}
    \caption{Mg II index composite created using the two synthetic timeseries shown in \fref{fig:mglong}b. For the interval 1745--1998, the facular area input from \citet{Wuetal2018} is used, while during 1999--present the facular area coverages based on \citet{Yeoetal2014} are used. The data used for this figure is available for download as supplementary material.}
    \label{fig:mgcomposite}
\end{figure*}

For comparison, in \fref{fig:mglong}a we also show the timeseries from \sref{sec:resobs} (red solid line) which gives an excellent agreement with the observations (shown in black). This figure implies that although the cycle amplitudes are comparable in the two versions, the Mg II index reconstructed from the \citet{Wuetal2018} facular coverages is generally lower than that based on the coverages from \citet{Yeoetal2014}. Since facular coverages of \citet{Yeoetal2014} are directly based on observed magnetograms, they are more accurate than coverages from \citet{Wuetal2018} that are based on sunspot number counts and model of the solar magnetic field evolution. Therefore, we carry out a linear regression and calibrate the Mg II index timeseries obtained using facular areas from \citet{Wuetal2018} to the same scale as the Mg II index timeseries obtained using faculae areas based on \citet{Yeoetal2014} as 

\begin{equation}
    {\rm Mg\ II}_{\rm Yeo\ et\ al.\ (2014)} = 1.113*{\rm Mg\ II}_{\rm Wu\ et\ al.\ (2018)} - 0.028.
\end{equation}
The result of this calibration is shown in \fref{fig:mglong}b. Finally, we obtain the Mg II index composite shown in \fref{fig:mgcomposite}, covering nearly three centuries. This composite is a combination of the \citet{Wuetal2018} based timeseries for the period 1745--1998 and the \citet{Yeoetal2014} based timeseries after 1998.

\section{Conclusions}
\label{sec:concl}
The Mg II index is a widely used proxy for solar UV irradiance. In this work, we have presented a model for the Mg II index variability based on the SATIRE approach. Specifically, we combined partial redistribution non-LTE radiative transfer computations of the Mg II line profiles in standard model atmosphere with magnetic filling factors taken from magnetograms. We have demonstrated that our model successfully reproduces the Mg II index variability observed by SORCE/SOLSTICE on activity cycle and on solar rotation timescales. We have created timeseries of the Mg II index spanning nearly three centuries, which is made available as a supplementary material. This timeseries could be used as an input for modeling the impact of solar variability, in particular in the UV, on Earth's climate. Furthermore, this timeseries could serve as an input in planetary atmosphere modeling approaches which use proxies of solar UV irradiance \citep[e.g.][see also \citealt{deOliveiraetal2024}]{Lundinetal2013,Peteretal2014}. 

%\begin{acknowledgements}
We thank the anonymous referee for their helpful comments. This study has made use of SAO/NASA Astrophysics Data System's bibliographic services and the open source Python package Matplotlib \citep{Hunter2007}. We acknowledge support from the European Research Council (ERC) under the European Union’s Horizon 2020 research and innovation program (grant No. 101118581 and No. 101097844) and the South African National Research Foundation (Grant Number 132800).
%\end{acknowledgements}

\bibliographystyle{aasjournal}
\bibliography{mgindex}

\end{document}